\documentclass[aps,pra,preprint,longbibliography,superscriptaddress,11pt]{revtex4-1}

\usepackage{color}
\usepackage{xcolor}
\usepackage{graphicx}
\usepackage{amsmath}
\usepackage{amsfonts}
\usepackage{soul}
\usepackage[normalem]{ulem}
\usepackage{lipsum}
\usepackage{bibentry}
\usepackage{float}
\usepackage{bm}

\usepackage{silence}
\nobibliography*

\begin{document}

\title{Analysis of interaction dynamics and rogue wave localization in modulation instability using data-driven dominant balance}

\author{Andrei V. Ermolaev}
\affiliation{Universit\'{e} de Franche-Comt\'{e}, Institut FEMTO-ST, CNRS UMR 6174, 25000 Besan\c{c}on, France}

\author{Mehdi Mabed}
\affiliation{Universit\'{e} de Franche-Comt\'{e}, Institut FEMTO-ST, CNRS UMR 6174, 25000 Besan\c{c}on, France}

\author{Christophe Finot}
\affiliation{Universit\'{e} de Bourgogne, Laboratoire Interdisciplinaire Carnot de Bourgogne, CNRS UMR 6303, 21078 Dijon, France}

\author{Go\"{e}ry Genty}
\affiliation{Photonics Laboratory, Tampere University, FI-33104 Tampere, Finland}

\author{John M. Dudley$^{\ast,}$}
\affiliation{Universit\'{e} de Franche-Comt\'{e}, Institut FEMTO-ST, CNRS UMR 6174, 25000 Besan\c{c}on, France}

\date{\today}

\vskip 2cm

\begin{abstract}
We analyze the dynamics of modulation instability in optical fiber (or any other nonlinear Schr\"{o}dinger equation system) using the machine-learning technique of data-driven dominant balance. We aim to automate the identification of which particular physical processes drive propagation in different regimes, a task usually performed using intuition and comparison with asymptotic limits. We first apply the method to interpret known analytic results describing Akhmediev breather, Kuznetsov-Ma, and Peregrine soliton (rogue wave) structures, and show how we can automatically distinguish regions of dominant nonlinear propagation from regions where nonlinearity and dispersion combine to drive the observed spatio-temporal localization.  Using numerical simulations, we then apply the technique to the more complex case of noise-driven spontaneous modulation instability, and show that we can readily isolate different regimes of dominant physical interactions, even within the dynamics of chaotic propagation.
\end{abstract}

\flushbottom
\maketitle

\section*{Introduction}

Modulation instability (MI) of the nonlinear Schr\"{o}dinger equation (NLSE) describes the process whereby a weak perturbation experiences exponential growth at the expense of a strong input wave \cite{Zakharov-1972,Zakharov-2009}. MI (sometimes called the Benjamin-Feir or Bespalov-Talanov instability \cite{Benjamin-1967,Bespalov-1966})  leads to complex spatio-temporal pattern formation, and is one of the fundamental nonlinear dynamical processes of nature. It has been observed in many different systems including hydrodynamics, plasmas, Bose-Einstein condensates, and fiber optics. Despite this large body of work over many years, its centrality to nonlinear science is such that it continues to be extensively studied from both experimental and theoretical perspectives.  Recent work, for example, has explored its description in terms of integrable turbulence \cite{Randoux-2016a,Walczak-2015}, its relationship with computational complexity \cite{Perego-2022}, its thermodynamic link to the soliton-gas concept \cite{Gelash-2019}, and its intrinsic association with Fermi-Pasta-Ulam-Tsingou recurrence \cite{Mussot-2018}; to cite only a small number of examples.

The dynamics of MI leads to the spontaneous emergence of localized structures that possess different spatial and/or temporal periodicities \cite{Akhmediev-1986,Akhmediev-1997}. These structures are intimately connected with known analytic solutions to the NLSE (including Peregrine soliton and Akhmediev and Kuznetsov-Ma breathers \cite{Kuznetsov-1977,Ma-1979,Peregrine-1983}), and understanding this correspondence has allowed experimentalists to excite a wide range of soliton and breather solutions in both optics and hydrodynamics \cite{Dudley-2014,Kibler-2021,Chabchoub-2013,Chabchoub-2011}. Moreover, even under conditions where modulation instability is excited from noise, it has been shown that the random peaks developing from noise possess the expected characteristics of these analytic solutions \cite{Dudley-2009,Toenger-2015,Narhi-2018}.  It is these nonlinear localization dynamics in particular that have attracted great interest as potentially underpinning the growth and decay of destructive rogue waves on the ocean \cite{Dudley-2019,Chen-2022}.

These various studies have yielded significant insights into the properties of MI under diverse conditions, and in different systems.  Somewhat surprisingly, however, although some aspects of MI localization can be interpreted precisely using mathematical methods such as the inverse scattering transform \cite{Randoux-2016b}, the physics of nonlinear and dispersive interactions in MI is more often discussed in qualitative terms by a comparison with specific limiting cases or characteristic nonlinear and dispersive length scales \cite{Agrawal-2019}. It would be highly desirable to have a means of interpreting the physics of MI that went beyond such a qualitative description, and yet avoided the formalism of the inverse scattering method.

In this paper, we show that the machine-learning technique of data-driven dominant balance can address this problem.  Machine learning methods are currently of great interest in all areas of physics \cite{Jordan-2015,Brunton-2016, Brunton-2022}, and in the particular field of nonlinear optics, have been applied to the study of various NLSE propagation scenarios  \cite{Genty-2020,Salmela-2021,Ermolaev-2022,Mabed-2022}.  The technique of dominant balance aims to automatically determine the contributing dominant physical processes at each step of propagation. As a subset of unsupervised learning techniques, it has been successfully applied to interpret the physics of a number of nonlinear propagation scenarios in hydrodynamics, as well as the more challenging case of broadband supercontinuum generation \cite{Callaham-2021a}.

In this paper, we use a dominant balance approach to analyse modulation instability of the NLSE.  We first apply the method to interpret known analytic solutions for Akhmediev breather, Kuznetsov-Ma, and Peregrine soliton structures, and for these spatio-temporal dynamics, we show how we can distinguish background regions of dominant nonlinear propagation from regions where nonlinearity and dispersion interact to drive localization.  This is especially important in showing how dominant balance can provide complementary insights into the dynamics, because associating the nonlinear stage of evolution with the background may seem counter-intuitive as this is a region of low intensity.  Following these studies of analytic SFB solutions we then use numerical simulations to study the more complex propagation case of noise-driven chaotic MI, and find again that we can automatically identify these different regimes of physical interaction.

\section*{NLSE Solutions}
We consider MI occurring in the focusing NLSE which is written in normalised form as follows:
\begin{equation}
    i\frac{\partial \psi}{\partial \xi} + \frac{\partial^2 \psi}{\partial \tau^2} + |\psi|^2 \psi = 0.
\label{eq: NLSE}
\end{equation}
Here $\psi(\xi,\tau)$ is a field envelope evolving in distance $\xi$ and co-moving time $\tau$. Dimensionless variables $\xi$ and $\tau$ are related to the usual notation of nonlinear optics by $\xi = z/L_\mathrm{NL}$ and $\tau = t/\sqrt{L_{NL}|\beta_{2}|/2}$, where $L_\mathrm{NL} = (\gamma P_0)^{-1}$. Here $z$ and $t$ are dimensional distance and time, $P_{0}$ is power (usually that of the input continuous wave), and $\beta_2$ and $\gamma$ are the usual dimensional fiber group velocity dispersion and nonlinearity parameters respectively \cite{Agrawal-2019}. The field envelope $\psi(\xi, \tau)$ is normalized with respect to $P_{0}^{1/2}$.

The NLSE possesses a number of known analytic solutions \cite{Akhmediev-1997,Sulem-1999}. Those associated with MI are the solitons on finite background (SFB), that can be written in compact form as follows:
\begin{equation}
    \psi(\xi,\tau) = \bigg[ 1 + \frac{2(1-2a)\cosh{(b \xi)} + i b \sinh{(b \xi)}}{\sqrt{2a} \cos(\omega_{m} \tau) - \cosh(b \xi)}\bigg] \exp(i \xi),
\label{eq:SFB}
\end{equation}
The physical behaviour of the solution is determined by the single governing parameter $a$ through arguments  $b=[8a(1-2a)]^{1/2}$ and $\omega_{m}=[2(1-2a)]^{1/2}$.  When $a=1/2$, $\omega_{m} = b = 0$ and the solution is the limiting rational Peregrine soliton, double-localized in $\xi$ and $\tau$ \cite{Peregrine-1983}.  For $a<1/2$, $\omega_{m}$ and $b$ are real, and we obtain the $\tau$-periodic Akhmediev breather, with $\omega_{m}$ and $b$ taking on physical significance of a modulation frequency and exponential growth/decay rate respectively.  When $a>1/2$, $\omega_{m}$ and $b$ become imaginary, and we obtain the $\xi$-periodic Kuznetsov-Ma solution.  These various SFB structures are well known, and have been observed in a range of experiments since 2010 \cite{Kibler-2010,Kibler-2012,Frisquet-2013}.

\section*{Implementing the dominant balance technique}
In this section, we give a general overview of how the dominant balance technique and algorithm are applied to nonlinear propagation in the NLSE.  Further details and references are given in the Methods section.   The dominant balance technique aims to automate the process of identifying the key interacting physical processes associated with different spatio-temporal regions of evolution. The technique involves several steps. The first is to determine the evolution of the field $\psi(\xi,\tau)$, and this is straightforward here as we have access to the analytic result in Eq.~(2). However, as we see below for noise-driven MI, the evolution can also be obtained using numerical integration of the NLSE. Indeed, in the most general case, this could also involve analysis of experimental data when access to full field information is available \cite{Ryczkowski-2018,Tikan-2018}.

The second step analyses the evolution $\psi(\xi,\tau)$ in its associated ``equation space,'' where each coordinate axis corresponds to a physical process defined by one of the terms in the governing NLSE (see Methods). Specifically, for each point $(\xi,\tau)$, the NLSE terms $\{i\psi_{\xi},\psi_{\tau \tau}, \psi|\psi|^2\}$ are separately computed, and we search for a ``dominant balance'' regime where the NLSE is approximately satisfied by only a subset of terms (the other terms contributing only negligibly.) As shown in Ref.~\cite{Callaham-2021a}, machine learning tools can automate this search, using cluster detection (Gaussian mixture modelling) and sparse regularization to identify regions where different combinations of terms drive the dynamics. These are standard tools of unsupervised learning and optimization, and allow robust detection of clusters even when they overlap (see Methods) \cite{Bishop-2008,Brunton-2022}.  When different clusters are found to possess the same sparcity pattern (significantly reduced variance in the same directions of equation space), these are grouped together to form a particular candidate ``balance model.''  In the case of the NLSE with three possible interacting terms, this process has a simple geometric interpretation: two interacting terms will be associated with a cluster falling on a line in the three-dimensional equation space, three interacting terms will be associated with a cluster in a plane.

When the data is fully grouped into balance models, the final step is to re-map the clusters back onto the $(\xi,\tau)$ space for comparison with the standard evolution dynamics. Visually, we do this by segmenting the original domain using a color key describing each balance model.  In our analysis, we used the code package described in Ref.~\cite{Callaham-2021a}, and available at the online repository \cite{Callaham-2021b}.  We also note that since we are dealing with complex fields, we stacked real and imaginary components as input to allow grouping of regions of significant variance irrespective whether identified in the real or imaginary components \cite{Callaham-2021b}.

\section*{Results}

We first apply this technique to identify locally-dominant interactions during the evolution of the three classes of SFB described above. Figure 1 shows results for the Peregrine soliton. Specifically, Fig.~1(a-i) shows the spatio-temporal evolution $|\psi(\xi,\tau)|^2$ which reveals the expected double-localization. The results of the dominant balance procedure are shown in Figure 1(b). Here Fig.~1(b-i) plots the identified clusters in the three-dimensional space of the real parts of coordinates  $\{i\psi_{\xi},\psi_{\tau \tau}, \psi|\psi|^2\}$, whereas Fig.~1(b-ii) and Fig.~1(b-iii) show two projections as indicated. The color key corresponds to two different dominant balance models that are found: one where only the nonlinear and propagation terms contribute (blue) and another where all NLSE terms contribute (orange). No cluster is found that involves only the dispersive and propagation terms. Note that for convenience we plot dependencies only for the real field components, but similar results are found for the imaginary components. The results in Fig.~1(b) show that all the points assigned to the blue cluster (nonlinear and propagation terms) are strongly localised in the equation space forming a dense distribution that manifests nearly zero variation with respect to the $\psi_{\tau \tau}$ axis (see particularly Fig.~1(b-iii)). In contrast, the orange cluster (all terms) is distributed throughout the equation space with no reduced variance with respect to any of three axes. This illustrates the geometrical interpretation lying behind the dominant balance approach.

\begin{figure}[htb!]
\centering\includegraphics[width=14cm]{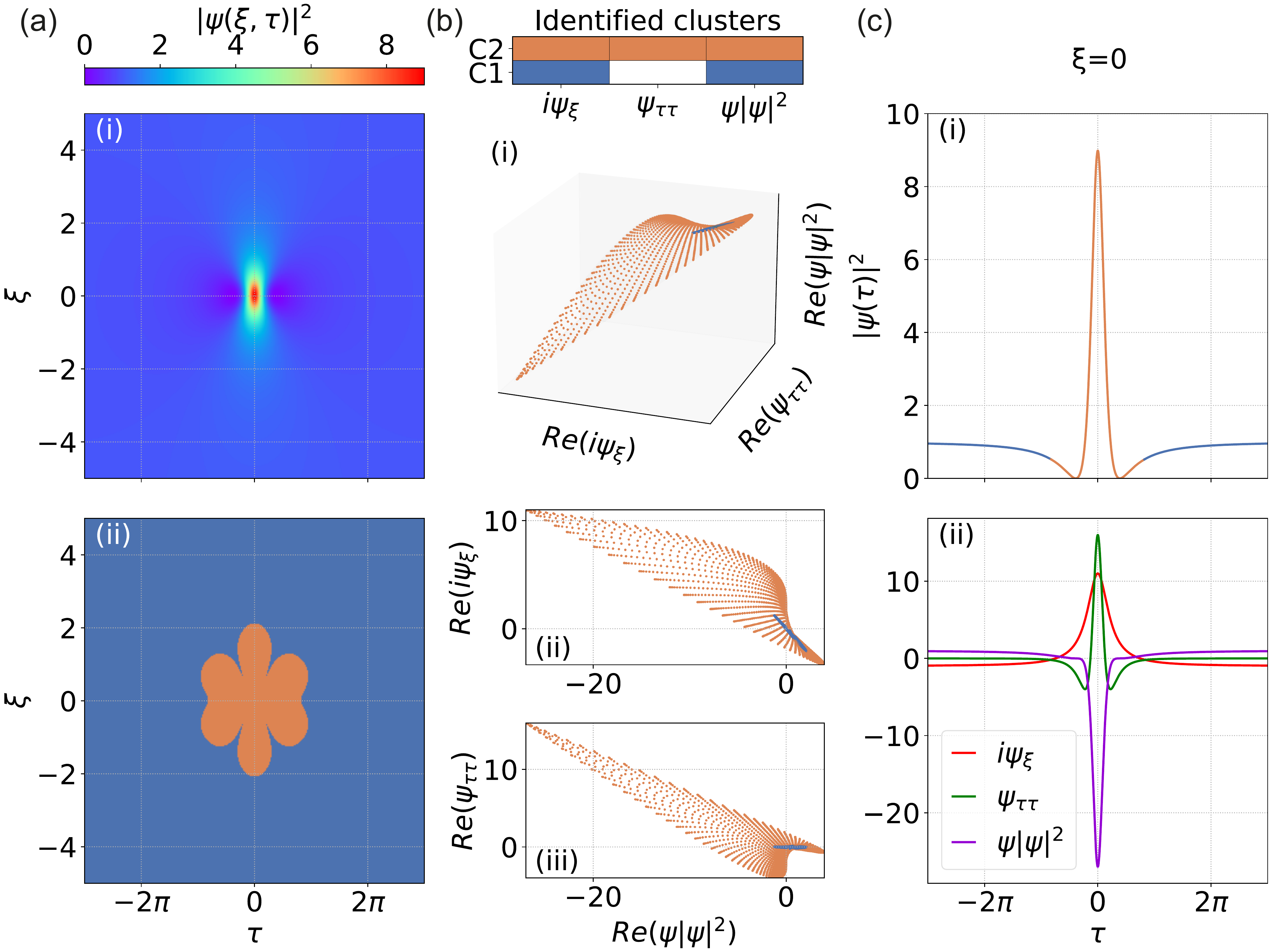}
\caption{\label{fig:1} Dominant balance method applied to the Peregrine soliton. (a-i) Spatio-temporal evolution of $|\psi(\xi,\tau)|^2$. (a-ii) Segmented map of the evolution space where the color key describes: only nonlinear and propagation terms (blue), and all NLSE terms (orange).  Using the same color key, (b) shows cluster identification for: (i) real parts of $\{i\psi_{\xi},\psi_{\tau \tau}, \psi|\psi|^2\}$; (ii) real parts of $\{\psi|\psi|^2 , i\psi_{\xi}\}$; (iii) real parts of $\{\psi|\psi|^2 , \psi_{\tau \tau}\}$. (c) Using the same color key, (i) shows the intensity profile at $\xi = 0$; (ii) Individual contributing terms in the NLSE at $\xi = 0$ as indicated in the legend.}
\end{figure}

The color-coded clusters are then mapped back onto a segmented dominant balance plot shown in Figure 1(a-ii), and the particular intensity profile at $\xi = 0$ is also plotted in Fig. 1(c-i) using the same color key. At $\xi = 0$, it is also instructive to plot the different contributions of each term of the equation space as shown in Fig.~1(c-ii), clearly revealing how different combinations of terms contribute to satisfy the NLSE (i.e. add to zero) in different regions. Note that at $\xi = 0$ all the three terms $\{i\psi_{\xi},\psi_{\tau \tau}, \psi|\psi|^2\}$ corresponding to the SFB solution are purely real.

These results reveal the key physical features of NLSE dynamics. For example, considering the Peregrine soliton and comparing Figs.~1(a-i) and 1(a-ii), the orange region reveals how the strong spatio-temporal localization around $(\xi = 0,\tau = 0)$ arises from the interaction between all terms in the NLSE, as both nonlinearity and dispersion combine to drive spatio-temporal compression. In contrast, the surrounding background region (blue) is dominated only by nonlinear evolution, and whilst this might be considered counter-intuitive since the background is where the intensity is lowest, this result actually highlights how interpreting NLSE physics requires comparison of the relative contributions of dispersion and nonlinearity. Specifically, a plane wave with no $\tau$-structure can not ``experience'' dispersion, and thus it is only nonlinear self-focussing that can initially influence the evolution of the background. It is only after temporal structure develops from this nonlinear stage of evolution that dispersion and nonlinearity interact. In fact, this approach to visualizing the evolution very clearly illustrates the well-known ``nonlinear'' stage of the instability \cite{Sulem-1999,Zakharov-2013}.  The ability of the dominant balance analysis to identify this nonlinear stage explicitly (even though perhaps counter-intuitive from a naive perspective) is an example of how it can yield important insights into nonlinear evolution.

\begin{figure}[htb!]
\centering\includegraphics[width=14cm]{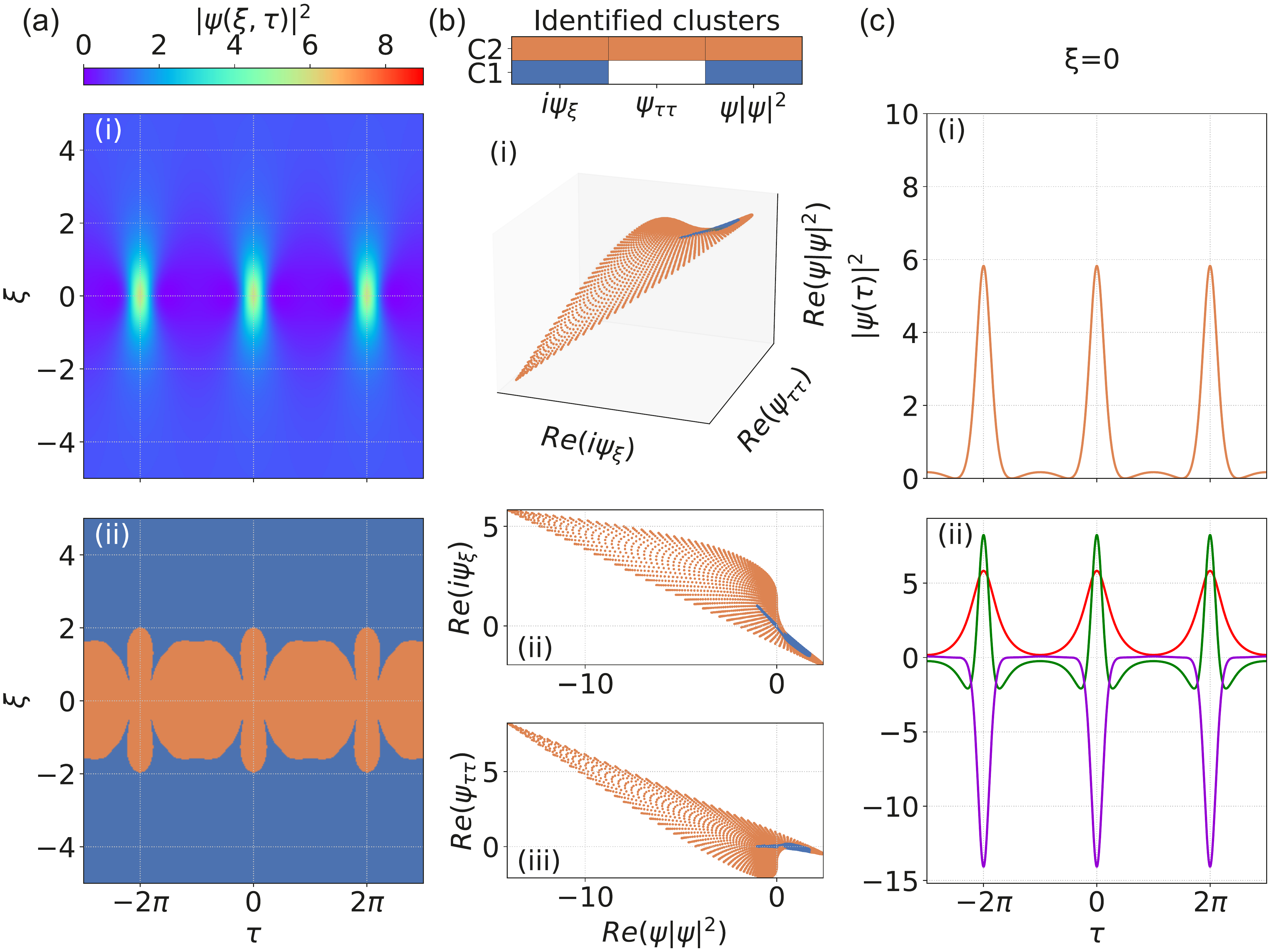}
\caption{\label{fig:2} Dominant balance method applied to the Akhmediev breather. (a-i) Spatio-temporal evolution of $|\psi(\xi,\tau)|^2$. (a-ii) Segmented map of the evolution space where the color key describes: only nonlinear and propagation terms (blue), and all NLSE terms (orange).  Using the same color key, (b) shows cluster identification for: (i) real parts of $\{i\psi_{\xi},\psi_{\tau \tau}, \psi|\psi|^2\}$; (ii) real parts of $\{\psi|\psi|^2 , i\psi_{\xi}\}$; (iii) real parts of $\{\psi|\psi|^2 , \psi_{\tau \tau}\}$. (c) Using the same color key, (i) shows the intensity profile at $\xi = 0$; (ii) Individual contributing terms in the NLSE at $\xi = 0$ as indicated in the legend of Fig.1(c-ii).}
\end{figure}

The results in Figs 2 and 3 for the Akhmediev and Kuznetsov-Ma breathers respectively have similar interpretation.  Here we see again see how regions of background associated only with dominant nonlinearity (blue) have been clearly identified, but we also clearly see how the contributions of all terms (orange) leads to the expected spatio-temporal localization characteristics. We also note how for the particular case of the Akhmediev breather, the $\xi = 0$ profile plot in Fig.~2(c) shows how all terms contribute to the dynamics in the lower amplitude regions between the localized peaks.  These analytical SFB solutions, of course, do not exhaust the full variety of localised structures appearing in MI such as higher-order solutions \cite{Akhmediev-2009}, breather or soliton collisions \cite{Agafontsev-2021}, ghost interactions \cite{Xu-2020} etc. However, these key examples provide a clear indication of how the dominant balance approach can complement existing techniques such as inverse scattering transform \cite{Sulem-1999,Randoux-2016b,Gelash-2022} in interpreting NLSE dynamics.

\begin{figure}[htb!]
\centering\includegraphics[width=14cm]{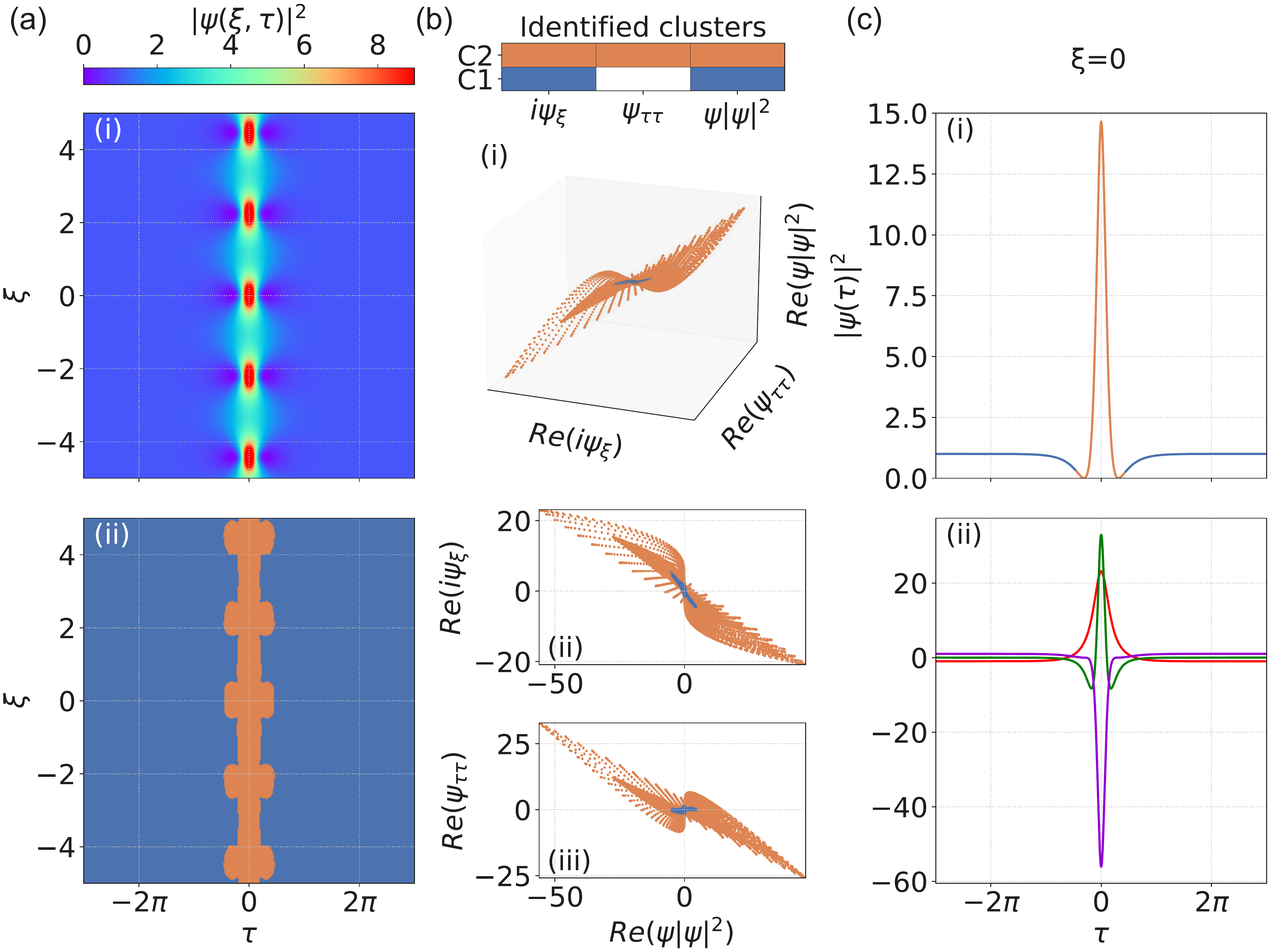}
\caption{\label{fig:3} Dominant balance method applied to the Kuznetsov-Ma breather. (a-i) Spatio-temporal evolution of $|\psi(\xi,\tau)|^2$. (a-ii) Segmented map of the evolution space where the color key describes: only nonlinear and propagation terms (blue), and all NLSE terms (orange).  Using the same color key, (b) shows cluster identification for: (i) real parts of $\{i\psi_{\xi},\psi_{\tau \tau}, \psi|\psi|^2\}$; (ii) real parts of $\{\psi|\psi|^2 , i\psi_{\xi}\}$; (iii) real parts of $\{\psi|\psi|^2 , \psi_{\tau \tau}\}$. (c) Using the same color key, (i) shows the intensity profile at $\xi = 0.$ (ii) Individual contributing terms in the NLSE at $\xi = 0$, as indicated in the legend of Fig.1(c-ii).}
\end{figure}

\begin{figure}[t!]
\centering\includegraphics[width=14cm]{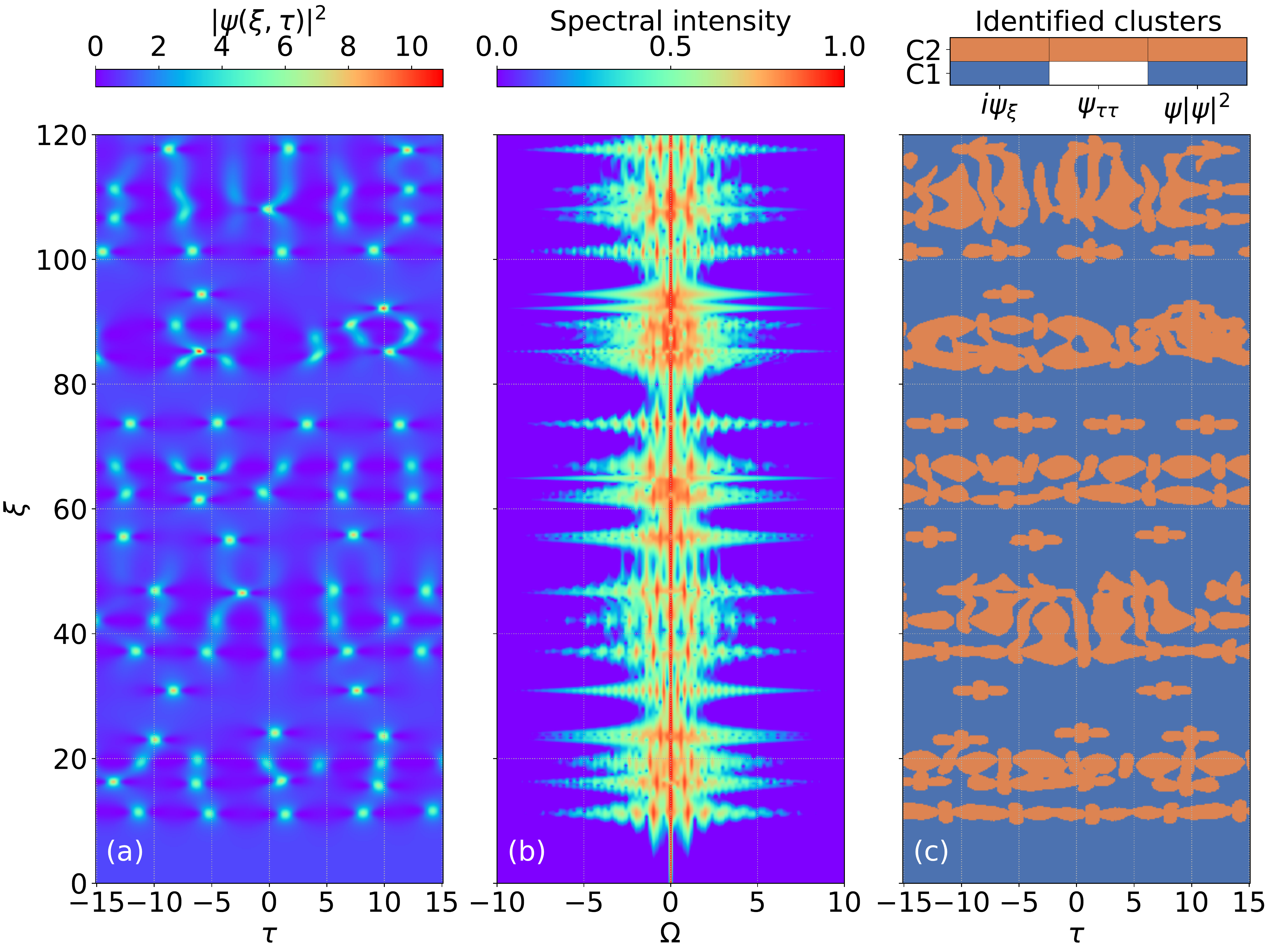}
\caption{\label{fig:4} (a) Spatio-temporal evolution of the normalized intensity for spontaneous modulation instability.  (b) The associated spectral evolution. (c) The results of dominant balance revealing the different interaction regions according to the colormap shown (same as in previous figures). }
\end{figure}

We now apply the dominant balance approach to interpret the more complex dynamics of noise-driven MI. For this case, the NLSE is solved numerically for a plane wave input with an imposed low level broadband noise background.  We used a common optical noise model corresponding to a one photon per mode background \cite{Dudley-2006}, but in fact similar chaotic dynamics in MI can be seen with essentially any class of random amplitude and/or phase fluctuation on the input \cite{Agrawal-2019}. The spatio-temporal intensity dynamics of $|\psi(\xi,\tau)|^2$ for this case are shown in Fig.~4(a) and for completeness, we also show in Fig.~4(b) the associated spectral evolution \cite{Dudley-2009}.

We clearly see how the input plane wave evolves into a series of localized peaks, displaying both random temporal (transverse) and spatial (longitudinal) periodicity. Maximum gain for the spontaneous instability is at sideband frequency $\Omega = 1$ and is associated with the initial emergence of Akhmediev breathers with temporal periodicity of $\Delta \tau = 2 \pi $. After this initial stage, subsequent evolution is plotted up to $\xi = 120$. We also see how the incoherent temporal evolution is reflected in the frequency domain with chaotic spectral expansion and contraction, as the random emergence of particularly high intensity temporal peaks of ultrashort duration is associated with broader spectra.

Analyzing the evolution in terms of dominant balance yields the results shown in Fig. 4(c). The color scale is the same as the previous figures. Comparing these results with the analytic SFB structures above allows us to distinguish the emerging localised structures. Indeed, even in this case of highly random MI dynamics data-driven dominant balance successfully finds the Akhmediev breathers with period $\Delta \tau \approx 2 \pi $ (for example, at $\xi \approx 11$ and $\xi \approx 38$), and we also see how propagation is associated with various $\xi$-periodic structures, breather collisions and Peregrine soliton-like rogue wave structures (e.g. the isolated feature in Fig. 4(c) at $\xi \approx 93$). Being based on unsupervised clustering of contributing terms to the evolution equation rather then simple intensity thresholding, the technique successfully identifies developing localised structures even in low-intensity regions. This suggests  the further application of the method in automated identification of emerging rogue wave structures \cite{Zou-2022}.

\section*{Discussion and Conclusions}

In conclusion, these results have clearly shown how the dominant balance approach provides a powerful tool for studying the interactions between dispersion and nonlinearity in the context of breather and modulation instability dynamics. In particular, even though these processes have been the subject of much previous study, visualising the dynamics with dominant balance segmentation clearly provides valuable insights into the relative contributions of different physical processes at different points in the evolution map.

We stress here, however, that data driven methods are not designed to replace existing techniques of analysing nonlinear dynamics, but should be seen as complementary tools to assist the use of physical considerations. For example, of particular interest is the way in which the dominant balance technique correctly associates the evolution of the plane wave background with a nonlinear stage of propagation. This illustrates how simplistic interpretations such as associating nonlinear evolution with intensity thresholding could be misleading, and it is always necessary to consider the relative contributions of nonlinearity and dispersion in discussing the dynamics of the NLSE.

Moreover, whilst with experience, inspection of spectral and temporal evolution maps of the NLSE can allow some processes (such as collisions) to be readily associated with combined nonlinear and dispersive interactions, such interpretations can sometimes be misleading. This is particularly the case with generalized forms of the NLSE where multiple processes combine, as previous studied  in Ref.~\cite{Callaham-2021a} for the case of optical fibre supercontinuum generation. The strength of the dominant balance approach is that it provides additional information in an unsupervised manner (i.e. not based on intuition or experience). When applied in parallel with other analysis techniques, this provides important complementary information to yield the best possible physical interpretation of complex evolution.

Finally, we note that the NLSE describes  propagation in many systems other than optical fiber, and there has been a strong recent focus on studying novel NLSE dynamics in deep-water hydrodynamics \cite{Dudley-2019}.  In this context, we anticipate an important area of future application will be the case of MI induced by localized perturbations \cite{Gelash-2014}, and the associated emergence of rogue wave statistics \cite{Kraych-2019,Gelash-2021}. There is clearly much potential for data-driven discovery methods to be applied in NLSE-related systems.\cite{Agrawal-2019}.

\newpage

\section*{Methods}

\noindent \textbf{Equation space representation} \\
The methodology of identifying a dominant balance model for a physical system at a particular stage of propagation aims to find a subset of terms of a more broadly applicable propagation model that locally dominates the dynamics.  Following the approach and notation of Ref.~\cite{Callaham-2021a}, we consider a general evolution equation on a domain $(\xi,\tau)$ written as follows:
\begin{equation}
     \sum_{i=1}^{K} f_{i}(\psi, \psi_{\xi}, \psi_{\tau} ..., \psi^{2}, \psi\psi_{\xi}, \psi\psi_{\tau},..., \psi_{\xi\xi}, \psi_{\tau\tau}, ... ) = 0,
\label{eq: general form}
\end{equation}
where $K$ is the number of terms, and the terms $f_{i}$ can be constructed in various ways from the spatio-temporal field $\psi(\xi, \tau)$. As discussed in Ref.~\cite{Callaham-2021a} (and its accompanying Supplementary Information), the advantage of this implicit form of the propagation equation is that it stresses the balance that must be present to satisfy the equality: the sum of all the terms must be zero. ``Dominant balance'' describes the situation when only a subset $p$ of the $K$ terms dominate the equality such that the contributions from the other $K-p$ terms are small or negligible.  Geometrically, the equation space is described by a vector: $\mathbf{f}(\xi,\tau) = [f_{1}[\psi(\xi_{n}, \tau_{m}),...], ...,f_{K}[\psi(\xi_{n}, \tau_{m}),...]]^{T}$ where each of the dimensions (directions) corresponds to a specific term in the evolution equation (here indices $n \in [1,N]$ and $m \in [1,M]$ represent the discretization of $\psi(\xi, \tau)$, where $N$ and $M$ are the number of points in the $\xi$ and $\tau$ directions respectively). A dominant balance regime then has a direct geometrical interpretation - dynamical points attributed to a certain dominant balance regime will be restricted to $p$ directions of the full $K$-dimensional space. In other words, when plotting the different terms in the equation space, the points associated with the dominant $p$ terms will have significantly reduced variance with respect to other $K-p$ directions.

In the case of the NLSE, the dimensionality $K=3$ and each dynamical point $\psi(\xi_{n}, \tau_{m})$ is associated with a vector $[i\psi_{\xi}(\xi_{n}, \tau_{m}),\psi_{\tau \tau}(\xi_{n}, \tau_{m}), |\psi(\xi_{n}, \tau_{m})|^{2}\psi(\xi_{n}, \tau_{m})]^{T}$.
In geometrical terms, dominant balance between the propagation and nonlinear Kerr terms $(i\psi_{\xi},|\psi|^{2} \psi)$ will be represented by an ensemble of points restricted on a line with near-zero variance with respect to the dispersion term $\psi_{\tau \tau}$ (e.g. the blue clusters in Figs. 1-3).  In contrast, the ensemble of points distributed throughout the $i\psi_{\xi}+\psi_{\tau \tau}+|\psi|^{2} \psi = 0$ plane will represent the full dynamics that involves the interplay of all three dynamical terms (e.g. the orange clusters in Figs 1-3).
\vskip 5mm

\newpage
\noindent \textbf{Finding Dominant Balance models through clustering} \\
The search for dominant combinations of terms within a higher-dimensional equation space is an ideal problem for unsupervised clustering algorithms \cite{Bishop-2008,Brunton-2022}.  In particular, we use the algorithm and code package described in Ref.~\cite{Callaham-2021a} and Ref.~\cite{Callaham-2021b} respectively which are based on a probabilistic Gaussian Mixture Model (GMM) framework. GMM seeks to locate clustered subpopulations within an overall population of data, under the assumption that the data consists of a mixture of Gaussian distributions with specified weights, means and covariance matrices (usually denoted $\pi_k, \boldsymbol{\mu}_k,\boldsymbol{\Sigma}_k$ respectively, where $k$ is the cluster index). The covariance matrix here generalizes the usual variance of a one-dimensional Gaussian distribution to higher dimensions. In contrast to simpler techniques such as k-means associated with hard partitions between clusters, GMM describes membership of a clusters in a probabilistic sense, allowing the algorithm to fit and return clusters that overlap.  The GMM algorithm is based on the expectation-maximisation technique, a standard approach that is fully described in e.g. Ref.~\cite{Bishop-2008}. The particular GMM algorithm used here is \texttt{GaussianMixture} from the \texttt{scikit-learn} Python package \cite{Pedragosa-2011}, as implemented in Ref.~\cite{Callaham-2021b}.

A key motivation to use the GMM is that the covariance matrices can be interpreted physically to identify combinations of terms that dominate the dynamics.  In particular, clusters associated with directions (dimensions) with significant variance correspond to physical terms that contribute actively to the dynamics (see the discussion of the results in Figs 1-3 above). However, there are some important additional factors that need to be considered to apply this approach successfully. In particular, since the data points in the equation space may not actually approximate a mixture of Gaussian distributions, the algorithm will usually return a number of clusters greater that the number of physical balance regimes. As described in detail in Ref.~\cite{Callaham-2021a}, this problem can be overcome using Sparse Principal Component Analysis (Sparse PCA) which uses $l_{1}$-regularisation to determine a sparse approximation to the leading principal component of each cluster \cite{Zou-2006,Jolliffe-2016}.  In this case, when a particular cluster is associated with a dominant balance regime, it should be well described by the particular direction of its maximum variance.  Note that $l_{1}$-regularisation in this context is a standard approach in machine learning using the $l_{1}$ norm as the penalty in the PCA regression-optimization problem \cite{Zou-2006}.

There are two key parameters that need to be selected to ensure that the returned models correspond as accurately as possible to physical regimes of dominant balance. The first is the particular number of clusters used in the Gaussian Mixture Model. Although in principle we can already anticipate the maximum number of potential clusters based on the number of terms in the propagation model, it is usually advantageous to initially choose a greater number, as the  $l_{1}$-regularisation step will later group together clusters found to possess the same sparcity patterns (i.e. reduced variances in the same directions of equation space) \cite{Callaham-2021a}.  The second parameter is associated with the  sparse regularisation of the PCA that describes the tradeoff between accuracy and sparsity in the returned models. A procedure for this selection process is described in detail in the Supplementary information of Ref.~\cite{Callaham-2021a}, and is based on considering a returned Pareto-type curve that plots the residual error of the inactive terms (accuracy) against the regularization parameter (sparsity). It is generally straightforward to see from this plot the most suitable parameter to generate the returned balance model. The very last step of the algorithm involves re-mapping the sparse clusters back onto the original spatio-temporal domain, and it is at this point we can directly compare the initial field distribution with the identified cluster map (as in Figs 1-4).

It is useful to give further numerical details for our results.  For the three classes of soliton on finite background considered in Figs 1-3, the evolution maps $\psi(\xi, \tau)$ were computed over $(N \times M) = (501 \times 1024)$ in $\xi$ and $\tau$ respectively. For the noise-driven map considered in Fig.~4, evolution was computed over $(N \times M) = (5001 \times 1024)$ in $\xi$ and $\tau$ respectively. The GMM search was based on an initial selection of up to 5 clusters and the sparse regularisation parameter $\alpha$ (used in the Python function \texttt{SparsePCA} \cite{Callaham-2021b}) was in the range 50--100. We also note the computation time associated with the GMM clustering and SPCA analysis, which was typically 6 and 21 minutes respectively for solitons on finite background and noise-driven MI, running on a standard Windows PC with 3.00~GHz 6 MB cache double-core CPU.

\vskip 5mm

\section*{Acknowledgements} Funding: Academy of Finland (318082, 320165 Flagship PREIN, 333949); Centre National de la Recherche  Scientifique (MITI Evènements Rares 2022); Agence Nationale de la Recherche (ANR-15-IDEX-0003, ANR-17-EURE-0002, ANR-20-CE30-0004). We thank Daniel Brunner and Pierre Colman for stimulating discussions.

\section*{Disclosures} The authors declare no conflicts of interest.

\section*{Data Availability} The data underlying the results presented in this paper are available from the corresponding author J.M.D. upon reasonable request.

\section*{Author contributions statement}
Simulations and analysis were performed by A.V.E. with guidance from J.M.D.  All authors were jointly involved in validation and interpretation of the results obtained, and in the writing of the article. J.M.D. provided overall project supervision.

\newpage


\end{document}